\def\gapprox{{_>\atop{^\sim}}}

\def\cmmt{\rm {cm^{-2}}}

\def\s-1{\rm {s^{-1}}}
\def\twco{$^{12}$CO}

\def\etal {et al.}
\def\kms {\hbox{${\rm km\,s}^{-1}$}}
\documentclass{aa}
\usepackage{graphicx}
\begin{document}
\title{A Molecular Tidal Tail in the Medusa Minor Merger}
\author{S.~Aalto\inst{1}, S.~H\"uttemeister\inst{2}, A. G. Polatidis\inst{1}}
\offprints{S. Aalto}
\institute{
 Onsala Rymdobservatorium, Chalmers Tekniska H\"ogskola, S - 43992 Onsala, Sweden
\and
Astronomisches Institut der Universit\"{a}t Bochum,
 Universit\"atsstra\ss{}e 150,  D - 44780 Bochum, Germany}
\date{Received  / Accepted  }
\titlerunning{Molecular Tail in the Medusa}

%
\abstract{We have detected CO 1-0 emission along the tidal tail of the NGC~4194
(the Medusa) merger. It is the first CO detection in the optical tail
of a minor merger.  Emission is detected both in the centre of the
tail and at its tip.  The molecular mass in the 33$''$ Onsala 20m beam is
estimated to be $\gapprox 8.5 \times 10^7$ M$_{\odot}$ which is at least 4\% of
the total molecular mass measured so far in this system. 
We suggest that the emission is a molecular tidal tail which is
part of the extended structure of the main body, and that the molecular gas was
thrown out by the collision instead of having formed in
situ from condensing atomic material.
We find it unlikely that the emission is associated with a tidal dwarf galaxy
(even if the future formation of such an object is possible), but high resolution
HI, CO and optical observations are necessary to resolve the issue. 
The Medusa is very likely the result of an elliptical+spiral collison and our detection
supports the notion that molecular gas in minor
mergers can be found at great distances from the merger centre.
 }

\maketitle
\section{Introduction}

Although tidal tails are often very gas rich, the gas in most cases is in the form of
atomic (HI) gas. Searches for molecular hydrogen in the tails of mergers and interacting galaxies
were long unsuccessful (e.g. Smith and Higdon 1994). The molecular gas in a major disk-disk merger
becomes heavily concentrated towards the inner kpc in most cases (e.g. Barnes \& Hernquist 1996).
Thus, an extended CO component in a coalesced, major merger is not expected.
Recently, CO emission has, however,
been found in tidal dwarf galaxies (TDGs) formed in the extended tails of interacting
galaxies and mergers (e.g. Braine \etal\ 2000). 
These objects are kinematically distinct and often found at the tip of the tidal tail
(Duc \& Mirabel 1994). Molecular masses of
$10^7 - 10^8$ M$_{\odot}$ seem to be typical for TDGs. 
The molecular gas in these TDGs is suggested to form inside the HI clouds, and in
this scenario was not
expelled from the parent galaxy in molecular form (e.g. Braine \etal\ 2000). 
In the M~81 galaxy group, molecular condensations without 
an apparent stellar component are found in two places (Brouillet \etal\ 
1992; Heithausen \& Walter 2000). The latter object is extended over more
than $1'$, and could be a TDG in formation.

A collision between an elliptical and a less-massive spiral is a minor merger that leads to
the formation of regular shells. Dynamical simulations of elliptical+spiral (E/S) collisions suggest
that, if the gas clouds are treated as particles,
a significant amount of the molecular gas may follow the disk stars into the tidal features  - instead of 
becoming concentrated into the central kpc (e.g. Kojima and Noguchi 1997). Thus searching for
molecular gas in the tidal tails of minor mergers may be a fruitful approach. 
In this picture, gas could also be associated with the shells which are otherwise assumed to
be gas-free.
Indeed, Charmandaris \etal\ (2000) detect
CO emission in the shells of the nearby elliptical Centaurus~A --- which is believed to be
the result of an E/S merger.

The Medusa merger, NGC~4194, (Table 1; Figure 2)  
($L_{\rm IR} = 8.5 \times 10^{10}$ L$_{\odot}$ at $D$=39 Mpc) belongs to a class of lower luminosity
mergers compared to Ultraluminous IR Galaxies, which have
$L_{\rm IR} \gapprox 10^{12}$ L$_{\odot}$. The Medusa has an extended region of
intense star formation (e.g. Prestwhich \etal\ 1994; Armus \etal\ 1990), a 
feature that might be common for many of these `intermediate luminosity' 
mergers. As the name indicates, the optical morphology of the Medusa merger
is spectacular with a knotty tidal tail stretching out 60$''$ (11.3 kpc) north of the main body.

We have studied the central starburst region of this merger in the CO 1-0 line with the OVRO array 
and the Onsala 20m telescope (see Aalto \& H\"uttemeister 2000 (AH) and references therein) and found a surprisingly
extended molecular cloud distribution (Figure 1). A significant amount of the CO flux resides in dust
lanes crossing the centre and curving into the base of the tidal tail. The morphology of the Medusa
combined with the extended gas distribution led us to suggest that it is the result of an E/S galaxy
collision (AH).

With the intention of investigating the notion that molecular clouds may follow the stars out
to great radii in minor mergers we have searched for CO 1--0 emission in the middle and tip
of the tidal tail of NGC 4194 with the
Onsala 20m telescope. We detect the presence of significant amounts of molecular gas in both
positions. The observations are presented in section two and a brief discussion of the
implications of the detection in section three.

\section{Observations and Results}

\begin{figure}
\centering 
\includegraphics[width=6cm]{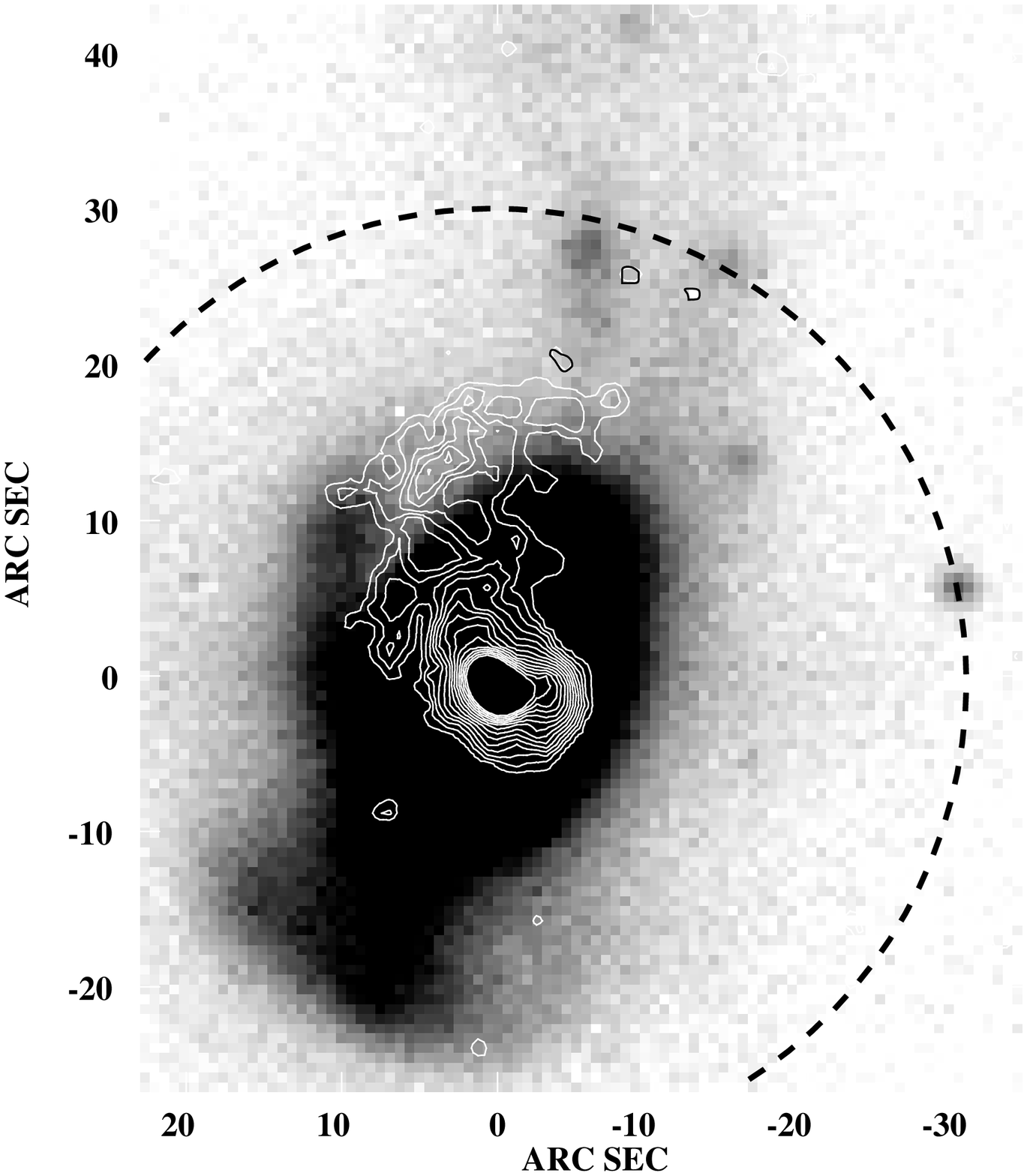}
\caption{Overlay of the OVRO CO (from AH) contours (white, apart from clouds in the
tail marked in black) on a greyscale, overexposed
optical R-band image (Mazzarella \& Boroson 1993). The dotted line shows the edge
of the OVRO primary beam. The contours are not primary beam corrected which means
that features close to the edge of the beam are suppressed in intensity.}

\label{something}
\end{figure}

\begin{table*}
\caption{\label{prop} Observed positions and Results$^{\rm a}$}
\begin{tabular}{lrrcccc}
Position & {R.A. (1950.0)} & {Dec (1950.0)} & $\int{T_{mb}}$ (K \kms) & $\Delta V$ (\kms)& $V_{\rm c}$ (\kms)
 & M(H$_2$)$^{\rm b}$ M$_{\odot}$\\
\hline \\
Centre & $12^{\rm h}\ 11^{\rm m}\ 41.22^{\rm s}$ & $54^{\circ}\ 48'\ 16''$ & $17 \pm 2$ & 200  
& 2530  & $\leq  2 \times 10^9$\\
Tail-1 & $12^{\rm h}\ 11^{\rm m}\ 40.00^{\rm s}$ & $54^{\circ}\ 48'\ 56''$ & $0.72 \pm 0.12$ 
& 100  & 2450  & $\geq 8.5 \times 10^7$\\
Tail-2 & $12^{\rm h}\ 11^{\rm m}\ 40.00^{\rm s}$ & $54^{\circ}\ 49'\ 12''$ & $0.45 \pm 0.14$ 
& 56  & 2448  & $\geq 5.3 \times 10^7$\\

\hline \\
\end{tabular} \\
a): For $\eta_{\rm beam}$=0.5 ; $\theta = 33''$\\
b): This is for a conversion factor of $X$ = N(H$_2$)/I(\twco)
= $2.3 \times 10^{20}$ $\cmmt$ (K \kms)$^{-1}$). The adopted distance
is $D$=39 Mpc.

\end{table*}

\begin{figure*}
\centering 
\includegraphics[width=17cm]{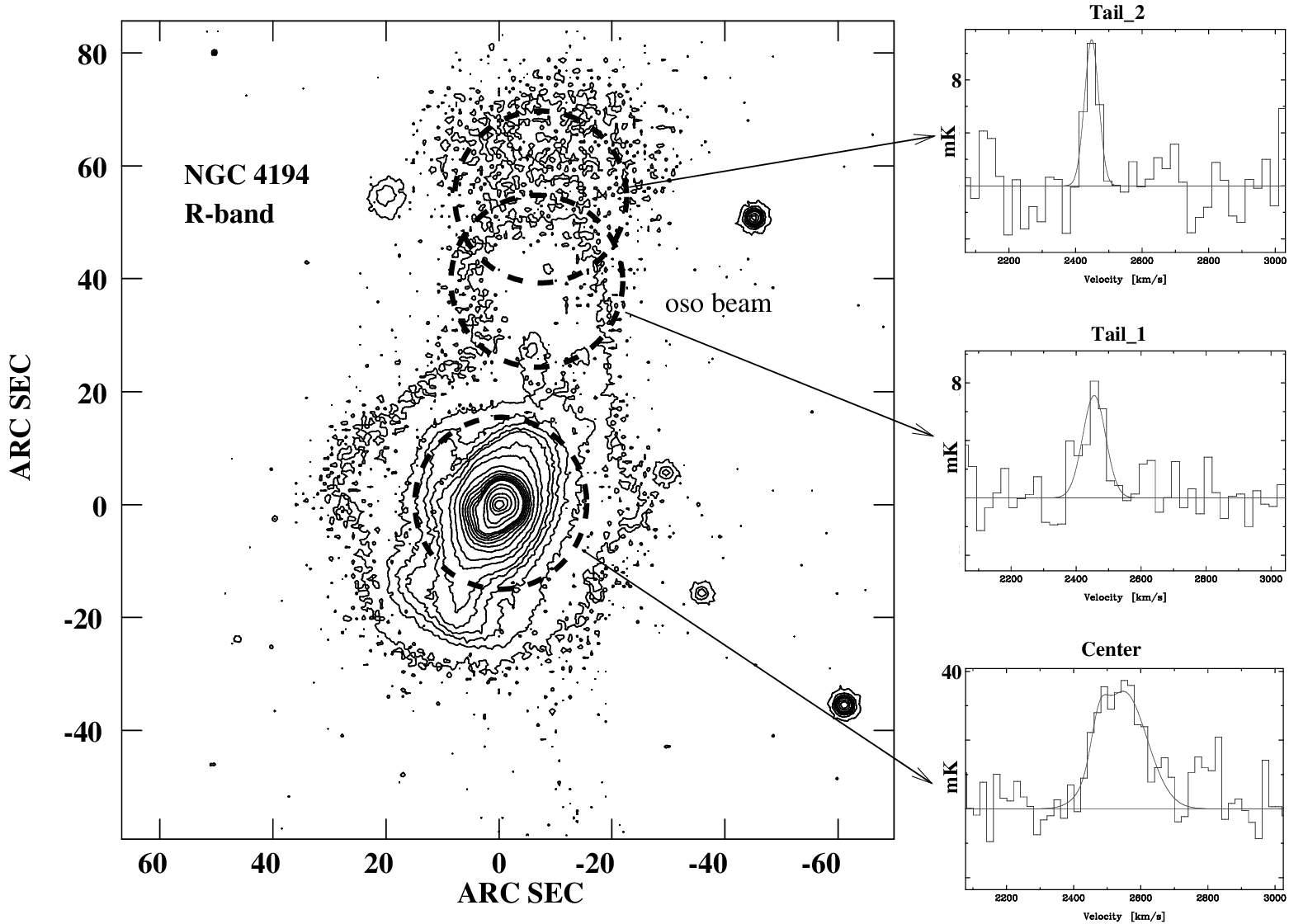}
\caption{To the left is a contour plot of an R-band CCD image of NGC~4194 
(Mazzarella and Boroson 1993). The levels are logarithmic (in flux) to clearly
show the faint structure, such as the northern tidal tail. The top and middle
right panels show the two CO 1--0 spectra in the tidal tail and the
bottom right panel shows the central spectrum. The temperature scale
is in mK (T$_{\rm mb}$).}

\label{something}
\end{figure*}

We have obtained single dish spectra of the 1--0 transition of CO in the tidal tail of 
NGC~4194 with the Onsala 20m telescope in two positions. The observations were carried out
in February 2001 and the 
system temperature was typically 500 K. We alternated between centre velocities
2300 \kms\ and 2500 \kms. The spectra were then joined together to get flat baselines
and minimize spurious spectrometer signals.
The pointing accuracy was checked on SiO masers and was found to be about 2$''$.
The observed positions, integrated intensities, central
velocities and molecular masses can be found in
Table 1. The signal in position 1 (mid-tail) is a $6\sigma$ detection and in position 2
(tip of tail) we are
detecting emission at the $3\sigma$ level, due to a higher rms noise and a
narrower linewidth. First order baselines have been removed in all
cases. For position 2, two bad channels were removed in the raw data at $v=2250$ \kms.
In Figure 2 we show the observed positions and spectra
overlayed on an R-band CCD-image (Mazzarella and Boroson 1993). The central spectrum is
taken from AH.

\section{Discussion}

\subsection{Molecular mass}
We have used a standard Galactic conversion factor to estimate the mass in both
the galaxy centre and in the tail. It is rather unlikely that the same conversion factor is
applicable for both regions since the centre emission originates in gas involved in
a starburst with associated extreme gas properties (AH).
We suspect that the standard conversion factor overestimates the molecular mass in
the centre, and underestimates it in the tail. It is difficult to judge by how large
a factor the conversion factor varies, but we believe that the gas detected in the
tail constitutes a larger fraction of the total H$_2$ mass than the 4\% indicated by
using the same factor. If the gas in the tail originated in the outskirts of the
precursor spiral, a conversion factor calibrated for the outer region of disk galaxies
should be appropriate (e.g. Arimoto \etal\ 1996).

The fact that the CO emission is detected far out in the actual optical tail sets
it apart from the tidal CO detection in the minor (disk/disk) merger NGC~2782
--- even if the linear distance from the centre of
the galaxy may be the same as that for our detection in NGC~4194.
In NGC~2782, $6 \times 10^8$ M$_{\odot}$
of molecular gas are associated with a depression in the optical emission at the base of the
eastern tail (Smith \etal\ 1999).

\subsection{Dynamics}
From our previous high-resolution OVRO map (AH) we see 
that the inner region of
NGC~4194 is rotating with a position angle of 160$^{\circ}$, where the northern
emission is blueshifted (even if the north-eastern part of the emission is
moving at too high velocity to fit in the simple rotational pattern). The CO emission
in the tidal tail is blueshifted by $\approx$100 
\kms\ from the centre gas (see Table 1) and thus appears to be participating in the general
rotational pattern of the main body.
In contrast, the tidal molecular material of NGC~2782 (Smith \etal\ 1999) is in
apparent counterrotation to the central gas  
which underlines the difference in merger history between NGC~4194 and NGC~2782.

\subsection{A molecular tidal tail?}

It is tempting to suggest that the tail emission is emerging from a TDG being born. 
The molecular mass indicated by the detection 
($> 8.5 \times 10^7 {\rm  M}_{\odot}$ for the solid detection in the middle
of the tidal tail) is compatible with the molecular
masses that have been reported to be associated with developed tidal dwarfs
in the Arp\,105 and Arp\,245 systems by Braine \etal\ (2000), which are 
$> 1.4 \times 10^8 {\rm  M}_{\odot}$ and $> 2.2 \times 10^8 {\rm  M}_{\odot}$,
respectively. However, we suggest that the
emission detected is still part of the structure in the main body and belongs to
the dust lane traced by CO emission in the northern
part of the main body (figure 2). The lane likely continues
well into the tidal tail, which is suggested by the blue image in Arp's
catalogue (1966). Thus, the CO emission detected indicates the presence of 
{\it a molecular tail}, perhaps as long as the optical tail.
We suggest that while the future destiny of the molecular emission may well be to
condense into a dwarf galaxy, this has not yet happened. However, high resolution
CO and HI observations together with sensitive optical studies are
necessary to determine whether the emission is coming from a distinct dynamical
system or not.

\subsection{The origin of the gas}

We suggest that the molecular gas in the tail of the Medusa was brought there
in molecular form instead of having formed in situ from atomic material. 
Although dust lanes and CO emission are not always related, in this particular galaxy there is
a very strong correlation (as shown by AH). A big fraction ($\gapprox$ 30\%) of
the CO emission is tracing dust lanes going from the centre all the way into
the tidal tail. We suggest
that, since the dust lane continues into the tidal tail, the CO continues
with it --- and that it is part of a molecular structure that followed
the stars into the tail, and remained molecular. One could of course argue that the CO-dust
correlation breaks down in the tail - but since we are detecting molecular gas in the
tail, it is natural to assume that it is associated with the dust.  

A single dish HI spectrum (Thuan and Martin 1981) shows the bulk of the HI emission
to be redshifted with respect to the tail. From the line profile,
we estimate that at most 1/3 of the total HI emission is associated
with the tail. However, to determine the origin of the molecular gas in the tail, both high resolution HI and CO maps are
necessary. For the H$_2$ to have formed from the HI we expect the CO emission to be associated with
column density peaks in the HI distribution. It could be argued that a more large-scale
phase transition between HI and H$_2$ has occured in which case a small scale relation between
HI and CO would not be expected. However, such phase transitions occur in high pressure
environments (such as in centres of galaxies) or in shocks. The tail of the Medusa appears to be
too diffuse to provide the right environment for such a transition.
Thus, differences in the CO and HI distribution should support the notion that the
H$_2$ did not form in situ. Multi-colour images in the optical will help study the structure of
the dust and its association to CO in more detail.

The situation in the Medusa is very different from major mergers where the CO emission
ends up collected in the inner region of the galaxy and where most of
the optical body of the merger is devoid of molecular gas. 
The CO found in full-blown optical TDGs of such major merger
systems (e.g. Braine \etal\ 2000) is very far from the merger centre and completely detached 
from it. Thus it may well have a different origin there, i.e. form from the
HI tails in situ. The connection between molecular gas in tidal tails
like that of the Medusa and TDGs clearly needs to be investigated further.

\subsection{The fate of the gas}

Armus \etal\ (1990) detect no H$\alpha$ emission in the
tail of the Medusa, so there is no indication of ongoing star formation in the molecular gas, even
if a deeper study will be necessary to determine actual star formation rates.
If the molecular gas in the tail of the Medusa is collected in self-gravitating clouds,
but quiescent,
it may survive for a longer time than the gas in the centre --- which is currently being
consumed by
a starburst. The gas clouds need to be part of a bigger structure, such as a
TDG, for them to survive when the tail disperses. 


\acknowledgements{
We thank J. Mazzarella for generously sharing his R-band image
of NGC~4194 with us. We thank the Onsala staff for their support 
during the observations. We also thank the referee, J. Braine,
for useful comments and suggestions.
This research has made use of the NASA/IPAC Extragalactic Database
(NED) which is operated by the Jet Propulsion Laboratory, California
Institute of Technology, under contract with the National Aeronautics and
Space Administration. }

\end{document}